\pdfoutput=1
\documentclass[aps,reprint,twocolumn,prd,nofootinbib,superscriptaddress]{revtex4-2}
\usepackage[utf8]{inputenc}
\usepackage{hyperref}
\hypersetup{colorlinks=true, citecolor=blue}
\usepackage{graphicx} 
\usepackage{amsmath,amssymb} 

\begin{document}
\title{Enhancing the significance of astrophysical events with multimessenger coincidences}

\author{Do\u{g}a Veske}
\email{veske@metu.edu.tr}
\affiliation{Institut f\"ur Theoretische Physik, Universit\"at Heidelberg, Heidelberg 69120, Germany}
\affiliation{Columbia Astrophysics Laboratory, Columbia University in the City of New York, New York, NY 10027, USA}
\affiliation{Fizik B\"ol\"um\"u, Orta Do\u{g}u Teknik \"Universitesi, \c{C}ankaya/Ankara 06800, Turkey}
\author{Albert Zhang}
\affiliation{Department of Physics, Columbia University in the City of New York, New York, NY 10027, USA}
\author{Zsuzsa M\'arka}
\affiliation{Columbia Astrophysics Laboratory, Columbia University in the City of New York, New York, NY 10027, USA}
\author{Szabolcs M\'arka}
\affiliation{Department of Physics, Columbia University in the City of New York, New York, NY 10027, USA}
\begin{abstract}

Coincident multimessenger observations of cosmic sources can offer numerous benefits, especially when used in the context of synergistic astrophysics. One significant advantage is enhancing the detection significance of separate detectors by correlating their data and assuming joint emission. We have formulated an approach for updating the Bayesian posterior probability of an astrophysical origin, namely $p_{\rm astro}$, relying on multimessenger coincidences assuming an emission model. The description is applicable to any combination of messengers. We demonstrated the formalism for the gravitational waves and high-energy neutrinos case. Applying our method to the public data of candidate coincident high-energy neutrinos with subthreshold gravitational-wave triggers, we found that in the case of highly energetic neutrino coincidences, $p_{\rm astro}$ can increase from approximately $\sim 0.1$ to $\sim 0.9$. The amount of improvement depends on the assumed joint emission model. If models are trusted, the marked improvement makes subthreshold detections much more confident. Moreover, the model dependency can also be used to test the consistency of different models. This work is a crucial step toward the goal of uniting all detectors on equal footing into a statistically integrated, Earth-sized observatory for comprehensive multimessenger astrophysics.
    
\end{abstract}
\maketitle
\section{Introduction}
Multimessenger detections~\cite{doi:10.1146/annurev.aa.27.090189.003213,Abbott_2017,blazar} are powerful observations that can unveil extra information about astrophysical events which could not be retrieved otherwise. Synergistic detections can uncover source dynamics physics as different messengers carry complementary information about the processes leading to the cataclysmic event. In search for coincident or synergistic detections of multiple messengers, many archival and real-time follow-ups have been carried out~\cite{2013JCAP...06..008A,2014PhRvD..90j2002A,Aartsen_2017_2,Albert_2019,Hamburg_2020,Aartsen_2020,2022icrc.confE.950V,Piotrzkowski_2022,Pillas_2023,Abbasi_2023,fletcher2023jointfermigbmswiftbatanalysis}. In addition to the physical knowledge we learn, the coincident or synergistic detections also provide more precise observational guidance; for example, the location of the source can be constrained better on the sky by considering the intersection of the independent localizations from different observatories. The narrower area can be searched more efficiently for further follow-ups (e.g., \cite{2021ApJ...909..126K}). Real-time multimessenger searches are especially useful for this purpose~\cite{2019ICRC...36..930K}. Another benefit of having associations between the observations of different detectors and cosmic messengers is that an uncertain detection can become a more confident one if a counterpart is found~\cite{2021ApJ...908..216V,1993ApJ...417L..17K}. Consequently, such joint analyses can improve the capability and impact of detectors by increasing their detection count and science reach via the optimal and efficient use of existing detector hardware, maximizing the impact of past scientific investment.

In this paper, we show an exact method to quantify the improvements on the significances of the triggers of a detector using the data of another detector, by assuming a joint emission for their messengers (e.g., \cite{ford2019agn,ford2019multimessenger,Kimura_2017,Kimura_2018,2017ApJ...849..153F}) that is mathematically applicable to any messenger combinations, including gamma-ray bursts, optical transients, neutrinos, or other kilonovae~\citep{2019LRR....23....1M} associated fingerprints. Our method , illustrated through a neutrino and gravitational wave use case for focus, is based on updating the Bayesian posterior probability of a  detected event candidate being astrophysical. Such a quantity is especially useful for quantifying the significances of gravitational-wave (GW) detections from compact binary coalescences, such as kilonovae. It is called $p_{\rm astro}$ and used as a decision statistic to classify the events as confident or subthreshold~\cite{PhysRevD.109.022001,Nitz_2021,PhysRevD.106.043009,PhysRevX.13.041039}. Unlike false-alarm rates, $p_{\rm astro}$ accounts for the rates of both noise and astrophysical triggers. Since the third observing run of the LIGO~\cite{2015CQGra..32g4001L} and Virgo~\cite{2015CQGra..32b4001A} detectors (O3), the threshold to include triggers in the catalogs as confident events have been conventionally set to $p_{\rm astro}\geq 0.5$, by the LIGO-Virgo-KAGRA Collaboration as well as by individual groups~\cite{PhysRevD.109.022001,Nitz_2021,PhysRevD.106.043009,PhysRevX.13.041039}. To have concreteness and due to its present practical use, we demonstrate our method considering significant improvements on the $p_{\rm astro}$ values of GW triggers. Again, for concreteness, we will use high-energy neutrinos as the complementary messenger  in our proof-of-principle example, mainly due to the elegance, availability, and quality of the real-time search of coincident high-energy neutrinos with GW events including the subthreshold ones. Consequently, this demonstration will be directly applicable to an existing follow-up effort that is a major goal of the fields and also defines the mathematical way for other prominent messengers from gamma-rays to optical observations. 

We start by deriving the formalism in Sec. \ref{sec:method}. In Sec. \ref{sec:data} we apply our method to the public data of candidate high-energy neutrino coincidences of the IceCube Observatory~\cite{Aartsen_2017,Aartsen_2024} with the subthreshold GW triggers from the first half of the fourth observing run of the LIGO~\cite{2015CQGra..32g4001L}, Virgo~\cite{2015CQGra..32b4001A} and KAGRA~\cite{10.1093/ptep/ptaa125} detectors (O4a)~\cite{2020LRR....23....3A}. We conclude in Sec. \ref{sec:conc}.
\section{Method}
\label{sec:method}
Let us start by analyzing the original $p_{\rm astro}$. Throughout the article we will denote the probability densities with $P$ and the probabilities with $p$. Let us express the GW emitting source being truly astrophysical as $\mathcal{T}$ and it being a false alarm as $\mathcal{F}$. Denoting the GW data as $GW$, we  have
\begin{multline}
\label{eq:pastro}
p_{\rm astro}=p(\mathcal{T}|GW)=\frac{P(GW|\mathcal{T})p(\mathcal{T})}{P(GW)}\\=\frac{P(GW|\mathcal{T})p(\mathcal{T})}{P(GW|\mathcal{T})p(\mathcal{T})+P(GW|\mathcal{F})p(\mathcal{F})}
\end{multline}
where $P(GW|\mathcal{F})$ is proportional to the false-alarm rate of the event, and $p(\mathcal{F})$ is the fraction of false alarms in the given set of triggers, but remarkably their values are not important for our final result. Knowing $p_{\rm astro}$, we can compute the only remaining quantity $P(GW|\mathcal{T})$.
\begin{equation}
\label{eq:2}
    P(GW|\mathcal{T})=\frac{p_{\rm astro}p(\mathcal{F})P(GW|\mathcal{F})}{(1-p_{\rm astro})p(\mathcal{T})}
\end{equation}

Now, assuming the GW event also emits high-energy neutrinos, let us find the updated $p_{\rm astro}$. Denoting the additional multimessenger data with $\nu$, we have the probability of the source being astrophysical
\begin{multline}
p(\mathcal{T}|GW,\nu)=\frac{P(GW,\nu|\mathcal{T})p(\mathcal{T})}{P(GW,\nu)}\\= \frac{P(GW,\nu|\mathcal{T})p(\mathcal{T})}{P(GW,\nu|\mathcal{T})p(\mathcal{T})+P(GW,\nu|\mathcal{F})p(\mathcal{F})}\\=\frac{P(GW|\mathcal{T})P(\nu|GW,\mathcal{T})p(\mathcal{T})}{P(GW|\mathcal{T})P(\nu|GW,\mathcal{T})p(\mathcal{T})+P(GW|\mathcal{F})P(\nu|\mathcal{F})p(\mathcal{F})}\\
=\frac{P(GW|\mathcal{T})p(\mathcal{T})}{P(GW|\mathcal{T})p(\mathcal{T})+P(GW|\mathcal{F})\frac{P(\nu|\mathcal{F})}{P(\nu|GW,\mathcal{T})}p(\mathcal{F})}.
\end{multline}

Comparing with the original $p_{\rm astro}$ in Eq. \eqref{eq:pastro}, we have a correction to one term. Plugging in $P(GW|\mathcal{T})$ from Eq. \eqref{eq:2} we get the multimessenger $p_{\rm astro}$ which we label as 
 $\mathcal{P}_{\rm astro}^{\rm MM}$
\begin{equation}
  \mathcal{P}_{\rm astro}^{\rm MM}=p(\mathcal{T}|GW,\nu)=\left(1+\frac{1-p_{\rm astro}}{p_{\rm astro}}\frac{P(\nu|\mathcal{F})}{P(\nu|GW,\mathcal{T})}\right)^{-1}.
  \label{eq:elevated}
\end{equation}

It is striking that without knowing anything about the inner details or Bayesian priors of the original $p_{\rm astro}$, nor additional information about the GW trigger, we can update its probability of being astrophysical with the data of additional messengers. \textit{Assuming} a joint emission, this extra likelihood ratio term in the expression decreases for ``better'' coincidences, which increases the $p_{\rm astro}$. 

Until now, $\nu$ has been a placeholder for the secondary messenger and one can even consider it as a set of several additional messengers. The structure of the likelihood ratio term will depend on how many and what kind of messengers one has. From this point on, we continue by considering the IceCube's high-energy neutrino data and expand these likelihoods as an example application  albeit any other messenger or messenger combinations can be used analogously.

\subsection{High-energy neutrino likelihood ratio}
For computing the likelihoods, we consider the following neutrino properties: their detection times ($t$), their energy proxies ($\epsilon$), which are related to the deposited energy to the detector, and their sky localizations. The sky localizations of the triggers arising from muon tracks can be approximated as 2-dimensional normal distributions that can be defined with the mean position ($\mathbf{\Omega}$) and a variance ($\sigma$) \cite{BRAUN2008299}. 
\subsubsection{Background likelihood}
Let us start with the background likelihood. The detection of background neutrinos can be modeled with a Poisson point process where every neutrino trigger is independent of each other. Considering we have $N$ neutrinos in our dataset, we have
\begin{equation}
    P(\nu|\mathcal{F})=Poi(N,n_{\rm av})\prod_{i=1}^NP(\nu_i|\mathcal{F})
\end{equation}
where indices represent every individual neutrino, $Poi(a,b)=b^ae^{-b}/a!$ is the Poisson point process probability, and $n_{\rm av}$ is the average number of neutrinos expected from the background during the duration of our dataset ($T$). Individual likelihood can be written as
\begin{equation}
P(\nu_i|\mathcal{F})=P(t_i|\mathcal{F})P(\epsilon_i,\mathbf{\Omega}_i|t_i,\mathcal{F})P(\sigma_i|t_i,\epsilon_i,\mathbf{\Omega}_i,\mathcal{F}).
\end{equation}

Ignoring minor seasonal variations~\cite{seasonal}, we consider background neutrino triggers to be equally likely to happen at any time, so the likelihood of time is just a constant.
\begin{equation}
P(t_i|\mathcal{F})=T_{\rm obs}^{-1}
\end{equation}
Furthermore, we consider the remaining likelihoods in the detector frame to be independent of the detection time, thanks to the stability of the IceCube detector. The sky position and energy proxy likelihood can be empirically obtained using past data.
\begin{equation}
    P(\epsilon_i,\mathbf{\Omega}_i|\mathcal{F})=P_{\rm emp}(\epsilon_i,\mathbf{\Omega}_i)
\end{equation}
Thanks to the axisymmetric location of IceCube at the South Pole with respect to the daily motion of Earth, the only sky direction on which the likelihoods depend is the declination in equatorial coordinates. The likelihood of localization uncertainty depends on the energy proxy and mean sky location only. We leave it as the following, which will cancel with the same term in the signal likelihood.
\begin{equation}
P(\sigma_i|\epsilon_i,\mathbf{\Omega}_i,\mathcal{F})=P(\sigma_i|\epsilon_i,\mathbf{\Omega}_i)
\label{eq:sigmaF}
\end{equation}
\subsubsection{Signal likelihood}
Now we move on to the signal likelihood. We consider different number of signal neutrinos to be detected ($\mathbb{N}\nu_{\rm det}$) within our dataset.
\begin{multline}
P(\nu|GW,\mathcal{T})\\= \sum_{n=1}^N P(\nu|GW,\mathcal{T},\mathbb{N}\nu_{\rm det}=n)p(\mathbb{N}\nu_{\rm det}=n|GW,\mathcal{T})\\+P(\nu|GW,\mathcal{T},\mathbb{N}\nu_{\rm det}=0)p(\mathbb{N}\nu_{\rm det}=0|GW,\mathcal{T})
\end{multline}

The likelihood when no signal neutrinos are detected is the same as the background likelihood.
\begin{equation}
    P(\nu|GW,\mathcal{T},\mathbb{N}\nu_{\rm det}=0)=P(\nu|\mathcal{F})
\end{equation}
Consequently, when no neutrino is expected to be detected, i.e. $p(\mathbb{N}\nu_{\rm det}=0|GW,\mathcal{T})=1$, the likelihood ratio in Eq. \eqref{eq:elevated} becomes 1, and we recover the original $p_{\rm astro}$ as we should.

The probability of detecting a certain number of neutrinos depends on the emission model we assume. Such high-energy emissions are expected in jets. Hence, we model the emission with a beaming factor $f_{\rm b}$ and a total neutrino emission energy $E_{\rm \nu tot}$. We approximate the emission spectrum as a power-law with index -2, which could be more sophisticated depending on the exact details of the models. With this model, we can write the probability of detecting $n\geq 1$ neutrinos as
\begin{multline}
\label{eq:detprob}
p(\mathbb{N}\nu_{\rm det}=n|GW,\mathcal{T})=\int P_{\rm GW}(\mathbf{\Omega},r)  \omega\\ \times Poi\left(n,\int \frac{A_{\rm eff}(\mathbf{\Omega},\epsilon)}{4\pi  r^2\epsilon^2} \frac{E_{\rm \nu tot}f_{\rm b}\mu}{\ln(\epsilon_{\rm max}/\epsilon_{\rm min})}d\epsilon\right)drd\mathbf{\Omega}, n\geq 1
\end{multline}
There, $r$ is the luminosity distance to the source, $P_{\rm GW}(\mathbf{\Omega},r)$ is the 3-dimensional localization of the GW detection, and $A_{\rm eff}$ is the energy and sky-position dependent effective area of the neutrino detector. $\epsilon_{\rm max}$ and $\epsilon_{\rm min}$ are the upper and lower bounds on the energies of the neutrinos in the considered emission. $\mu$ is the fraction of emitted energy that we can detect, e.g., we will be interested only in muon neutrinos, which will carry a fraction of the emitted total energy. $\omega$ is the fraction of the GW detections, which will have a beamed emission pointed toward the Earth. If the GW detections were independent of source inclinations, $\omega$ would be equal to $f_{\rm b}^{-1}$. However, GW detections from compact binaries favor low orbital inclinations, which makes $\omega\geq f_{\rm b}^{-1}$. Correspondingly, the probability of detecting no signal neutrinos from the GW source is
\begin{multline}
p(\mathbb{N}\nu_{\rm det}=0|GW,\mathcal{T})=\int P_{\rm GW}(\mathbf{\Omega},r) \bigg\{1- \omega\\ \times \left[1-\exp\left(-\int \frac{A_{\rm eff}(\mathbf{\Omega},\epsilon)}{4\pi  r^2\epsilon^2} \frac{E_{\rm \nu tot}f_{\rm b}\mu}{\ln(\epsilon_{\rm max}/\epsilon_{\rm min})}d\epsilon\right)\right]\bigg\}drd\mathbf{\Omega}.
\end{multline}

Now, we examine the only remaining term $P(\nu|GW,\mathcal{T},\mathbb{N}\nu_{\rm det}=n)$. We expand it by considering all of the possible detected astrophysical neutrino combinations $\{\nu\}$ in our dataset that has $\mathbb{N}\nu_{\rm det}=n$.
\begin{multline}
P(\nu|GW,\mathcal{T},\mathbb{N}\nu_{\rm det}=n)\\=\sum_{\{\nu\}:|\{\nu\}|=n}  \big[P(\nu|GW,\mathcal{T},\mathbb{N}\nu_{\rm det}=n, \nu_{det}=\{\nu\})\\ \times p( \nu_{det}=\{\nu\}|GW,\mathcal{T},\mathbb{N}\nu_{\rm det}=n)\big]
\end{multline}

Since every combination is equally likely \emph{a priori}, we have 

\begin{equation}
p(\nu_{det}=\{\nu\}|GW,\mathcal{T},\mathbb{N}\nu_{\rm det}=n)={N \choose n}^{-1}.
\end{equation}

Finally, we examine the term $P(\nu|GW,\mathcal{T},\mathbb{N}\nu_{\rm det}=n, \nu_{det}=\{\nu\})$.
We first separate the background likelihood for the $N-n$ neutrinos and the signal likelihood for $n$ neutrinos.
\begin{multline}
P(\nu|GW,\mathcal{T},\mathbb{N}\nu_{\rm det}=n, \nu_{det}=\{\nu\})\\=\left(Poi(N-n,n_{\rm av})\prod_{{\nu_i \notin \{\nu\}}} P(\nu_i|\mathcal{F}) \right) \\ \times P(\{\nu\}|GW,\mathcal{T},\mathbb{N}\nu_{\rm det}=n, \nu_{det}=\{\nu\})
\end{multline}
The time dependence of the signal likelihood is again independent of the rest of the observables. 
\begin{multline}
 P(\{\nu\}|GW,\mathcal{T},\mathbb{N}\nu_{\rm det}=n, \nu_{det}=\{\nu\})\\=   P(\{t\}|GW,\mathcal{T},\mathbb{N}\nu_{\rm det}=n, \nu_{det}=\{\nu\})\\ \times P(\{\epsilon,\mathbf{\Omega},\sigma\}|GW,\mathcal{T},\mathbb{N}\nu_{\rm det}=n, \nu_{det}=\{\nu\})
\end{multline}
As the emission model, we assume the emission flux starts $\Delta t^-$ before the time of the GW emitting merger $t_{\rm m}$ and ends after a duration of $\Delta t^+$ after that. We take these values as $\Delta t^-=\Delta t^+=\Delta t= 500$~s along the time windows of the searches \cite{2011APh....35....1B} and assume that the expected emission flux linearly increases up to the merger time and then linearly decreases to zero. There is no robust physical argument for or against this assumption. Our rationale is having the simplest continuous behavior for having a time window. Furthermore, it can be motivated by having neutrino emission and merger times being uniformly and independently distributed around a common reference time. The convolution of two uniform distributions gives the triangle distribution we describe. Moving on, the emissions of neutrinos are also independent of each other since the emission is assumed to be a Poisson point process, so the temporal likelihood is
\begin{multline}
P(\{t\}|GW,\mathcal{T},\mathbb{N}\nu_{\rm det}=n, \nu_{det}=\{\nu\})\\ =\prod_{t_i \in \{t\}} \frac{1}{\Delta t }-\frac{|t_i-t_{\rm m}|}{\Delta t^2} ,\ |t_i-t_{\rm m}| <\Delta t .
\end{multline}

To write the likelihood of $\{\epsilon,\mathbf{\Omega},\sigma\}$ we expand it using the real sky position of the source ($\mathbf{\Omega}_s$).
\begin{multline}
P(\{\epsilon,\mathbf{\Omega},\sigma\}|GW,\mathcal{T},\mathbb{N}\nu_{\rm det}=n, \nu_{det}=\{\nu\})\\=\int P(\{\epsilon,\mathbf{\Omega},\sigma\}|\mathbf{\Omega}_s,GW,\mathcal{T},\mathbb{N}\nu_{\rm det}=n, \nu_{det}=\{\nu\})\\ \times P(\mathbf{\Omega}_s|GW,\mathcal{T})d\mathbf{\Omega}_s
\end{multline}

The term $P(\mathbf{\Omega}_s|GW,\mathcal{T})$ is the sky localization of the source from the GW data.
\begin{equation}
    P(\mathbf{\Omega}_s|GW,\mathcal{T})=\int P_{\rm GW}(\mathbf{\Omega}_s,r)dr=P_{\rm GW}(\mathbf{\Omega}_s)
\end{equation}

For the fixed sky position, the observables of the individual neutrinos are independent. We also drop the ``$\mathbb{N}\nu_{\rm det}=n, \nu_{det}=\{\nu\}$" from the notation of individual likelihoods.

\begin{multline}
    P(\{\epsilon,\mathbf{\Omega},\sigma\}|\mathbf{\Omega}_s,GW,\mathcal{T},\mathbb{N}\nu_{\rm det}=n, \nu_{det}=\{\nu\})\\=\prod_{i=1}^n P(\epsilon_i,\mathbf{\Omega}_i,\sigma_i|\mathbf{\Omega}_s,GW,\mathcal{T})
\end{multline}

We manipulate the individual likelihood, first using Bayes' rule and then separating some of the joint probabilities.
\begin{multline}
P(\epsilon_i,\mathbf{\Omega}_i,\sigma_i|\mathbf{\Omega}_s,GW,\mathcal{T})\\=\frac{P(\mathbf{\Omega}_s,GW|\epsilon_i,\mathbf{\Omega}_i,\sigma_i,\mathcal{T})P(\epsilon_i,\mathbf{\Omega}_i,\sigma_i|\mathcal{T})}{P(\mathbf{\Omega}_s,GW|\mathcal{T})} 
\\=\frac{P(\mathbf{\Omega}_s|\epsilon_i,\mathbf{\Omega}_i,\sigma_i,\mathcal{T})P(GW|\mathbf{\Omega}_s,\epsilon_i,\mathbf{\Omega}_i,\sigma_i,\mathcal{T})}{P(\mathbf{\Omega}_s|\mathcal{T})P(GW|\mathbf{\Omega}_s,\mathcal{T})}\\ \times P(\epsilon_i,\mathbf{\Omega}_i|\mathcal{T})P(\sigma_i|\epsilon_i,\mathbf{\Omega}_i,\mathcal{T})
\end{multline}
$P(GW|\mathbf{\Omega}_s,\epsilon_i,\mathbf{\Omega}_i,\sigma_i,\mathcal{T})$ and $P(GW|\mathbf{\Omega}_s,\mathcal{T})$ terms are the same and cancel each other as GW data does not depend on neutrino observables. $P(\mathbf{\Omega}_s|\mathcal{T})$ is constant over the sky. Neglecting the minor variations of the effective area of the neutrino detector within the sky localization uncertainty of neutrinos $(\sim 1~{\rm deg}^2)$, $P(\epsilon_i,\mathbf{\Omega}_i|\mathcal{T})$ is proportional to the emission spectrum times the effective area values of the energy proxy and the mean sky position up to a normalization constant.
\begin{equation}
    P(\epsilon_i,\mathbf{\Omega}_i|\mathcal{T}) \propto \epsilon_i^{-2}A_{\rm eff}(\mathbf{\Omega}_i,\epsilon_i)
\end{equation}

The sky localization uncertainty likelihood is the same as the corresponding term in the background likelihood and they cancel each other when we take the likelihood ratio, i.e. like Eq. \eqref{eq:sigmaF} $P(\sigma_i|\epsilon_i,\mathbf{\Omega}_i,\mathcal{T})=P(\sigma_i|\epsilon_i,\mathbf{\Omega}_i)$. $P(\mathbf{\Omega}_s|\epsilon_i,\mathbf{\Omega}_i,\sigma_i,\mathcal{T})$ is the neutrino localization as a 2-dimensional normal distribution.
\begin{equation}
P(\mathbf{\Omega}_s|\epsilon_i,\mathbf{\Omega}_i,\sigma_i,\mathcal{T})=P_{\nu}(\mathbf{\Omega}_s|\mathbf{\Omega}_i,\sigma_i)=\frac{\exp\left(-\frac{|\mathbf{\Omega}_s-\mathbf{\Omega}_i|}{2\sigma_i^2}\right)}{2\pi \sigma_i^2}
\end{equation}

Wrapping up, we have
\begin{widetext}
\begin{multline} P(\nu|GW,\mathcal{T},\mathbb{N}\nu_{\rm det}=n, \nu_{det}=\{\nu\}) = \left(Poi(N-n,n_{\rm av})\prod_{{\nu_i \notin \{\nu\}}} P(\nu_i|\mathcal{F}) \right) \\ \times\frac{\int P_{\rm GW}(\mathbf{\Omega}_s)\prod_{{\nu_i \in \{\nu\}}} \left[\epsilon_i^{-2}P_{\nu}(\mathbf{\Omega}_s|\mathbf{\Omega}_i,\sigma_i)A_{\rm eff}(\mathbf{\Omega}_i,\epsilon_i)\left(\frac{1}{\Delta t }-\frac{|t_i-t_{\rm m}|}{\Delta t^2}\right)P(\sigma_i|\epsilon_i,\mathbf{\Omega}_i)\right]d\mathbf{\Omega}_s}{{\rm Normalization\ of\ the\ numerator\ for\ \epsilon_i\ and\ \mathbf{\Omega}_i}}.
\end{multline}
\end{widetext}
\subsubsection{Likelihood ratio}
We can finally write the likelihood ratio. Considering the detection probability is even low for detecting a single neutrino, we write it assuming at most 2 neutrinos that are also coincident with each other (we consider $n=1,\ 2$ by assuming $p(\mathbb{N}\nu_{\rm det}>2|GW,\mathcal{T})\approx0$). The likelihood ratio becomes
\begin{widetext}
\begin{multline}
\frac{P(\nu|GW,\mathcal{T})}{P(\nu|\mathcal{F})}
=\frac{Poi(N-1,n_{\rm av})}{Poi(N,n_{\rm av})}
\times \int P_{\rm GW}(\mathbf{\Omega},r)  \omega Poi\left(1,\int \frac{A_{\rm eff}(\mathbf{\Omega},\epsilon)}{4\pi  r^2\epsilon^2} \frac{E_{\rm \nu tot}f_{\rm b}\mu}{\ln(\epsilon_{\rm max}/\epsilon_{\rm min})}d\epsilon\right)drd\mathbf{\Omega}
\\ \times N^{-1}\sum_{i=1}^N P_{\rm emp}(\epsilon_i,\mathbf{\Omega}_i)^{-1}\left(\frac{T_{\rm obs}}{\Delta t}-\frac{T_{\rm obs}|t_m-t_i|}{\Delta t ^2}\right)\frac{\epsilon_i^{-2}A_{\rm eff}(\mathbf{\Omega}_i,\epsilon_i)\int P_{\rm GW}(\mathbf{\Omega})P_{\nu}(\mathbf{\Omega}|\mathbf{\Omega}_i,\sigma_i)d\mathbf{\Omega}}{\int [\epsilon_\nu'^{-2}A_{\rm eff}(\mathbf{\Omega}_\nu',\epsilon_\nu')\int P_{\rm GW}(\mathbf{\Omega})P_{\nu}(\mathbf{\Omega}|\mathbf{\Omega}_\nu',\sigma_i)d\mathbf{\Omega}]d\mathbf{\Omega}_\nu'd\epsilon_\nu'} 
\\
+\frac{Poi(N-2,n_{\rm av})}{Poi(N,n_{\rm av})}
\times \int P_{\rm GW}(\mathbf{\Omega},r)  \omega Poi\left(2,\int \frac{A_{\rm eff}(\mathbf{\Omega},\epsilon)}{4\pi  r^2\epsilon^2} \frac{E_{\rm \nu tot}f_{\rm b}\mu}{\ln(\epsilon_{\rm max}/\epsilon_{\rm min})}d\epsilon\right)drd\mathbf{\Omega}
\\ \times {N \choose 2}^{-1}\sum_{i=1}^{N-1} \sum_{j=i+1}^N \bigg[ P_{\rm emp}(\epsilon_i,\mathbf{\Omega}_i)^{-1}\left(\frac{T_{\rm obs}}{\Delta t}-\frac{T_{\rm obs}|t_m-t_i|}{\Delta t ^2}\right)P_{\rm emp}(\epsilon_j,\mathbf{\Omega}_j)^{-1}\left(\frac{T_{\rm obs}}{\Delta t}-\frac{T_{\rm obs}|t_m-t_j|}{\Delta t ^2}\right) \\ \times \frac{\epsilon_i^{-2}\epsilon_j^{-2}A_{\rm eff}(\mathbf{\Omega}_i,\epsilon_i)A_{\rm eff}(\mathbf{\Omega}_j,\epsilon_j)\int P_{\rm GW}(\mathbf{\Omega}) P_{\nu}(\mathbf{\Omega}|\mathbf{\Omega}_i,\sigma_i)P_{\nu}(\mathbf{\Omega}|\mathbf{\Omega}_j,\sigma_j)d\mathbf{\Omega}}{\int [\epsilon_\nu'^{-2}\epsilon_\nu''^{-2}A_{\rm eff}(\mathbf{\Omega}_\nu',\epsilon_\nu')A_{\rm eff}(\mathbf{\Omega}_\nu'',\epsilon_\nu'')\int P_{\rm GW}(\mathbf{\Omega})P_{\nu}(\mathbf{\Omega}|\mathbf{\Omega}_\nu',\sigma_i)P_{\nu}(\mathbf{\Omega}|\mathbf{\Omega}_\nu'',\sigma_j)d\mathbf{\Omega}]d\mathbf{\Omega}_\nu'd\epsilon_\nu'd\mathbf{\Omega}_\nu''d\epsilon_\nu''} \bigg]
\\
+\int P_{\rm GW}(\mathbf{\Omega},r) \left\{1- \omega\left[1-\exp\left(-\int \frac{A_{\rm eff}(\mathbf{\Omega},\epsilon)}{4\pi  r^2\epsilon^2} \frac{E_{\rm \nu tot}f_{\rm b}\mu}{\ln(\epsilon_{\rm max}/\epsilon_{\rm min})}d\epsilon\right)\right]\right\}drd\mathbf{\Omega}.
\end{multline}
\end{widetext}
There are some simplifications we can make
\begin{itemize}
    \item $\frac{Poi(N-1,n_{\rm av})}{Poi(N,n_{\rm av})}N^{-1}=1/n_{\rm av}, N\geq 1$
    \item $\frac{Poi(N-2,n_{\rm av})}{Poi(N,n_{\rm av})}{N \choose 2}^{-1}=2/n_{\rm av}^2, N\geq 2$
    \item $\int A_{\rm eff}(\mathbf{\Omega}_\nu,\epsilon)P_\nu(\mathbf{\Omega}|\mathbf{\Omega}_v,\sigma)d\mathbf{\Omega}_v\approx A_{\rm eff}(\mathbf{\Omega},\epsilon)$. This approximation is true for two reasons: First, $P_\nu(\mathbf{\Omega}|\mathbf{\Omega}_v,\sigma)$ is normalized such that $\int P_\nu(\mathbf{\Omega}|\mathbf{\Omega}_v,\sigma) d\mathbf{\Omega}=1$; but it is a symmetric distribution for $\mathbf{\Omega}$ and $\mathbf{\Omega}_\nu$. Hence $\int P_\nu(\mathbf{\Omega}|\mathbf{\Omega}_v,\sigma) d\mathbf{\Omega}_\nu=1$ is also true. Then, $\int A_{\rm eff}(\mathbf{\Omega}_\nu,\epsilon)P_\nu(\mathbf{\Omega}|\mathbf{\Omega}_v,\sigma)d\mathbf{\Omega}_v$ is just smearing out the effective area with the localization as the kernel. Second, the extent of localization ($\sim 1$~deg) is smaller than the scale of effective area's changes. Therefore that smearing does not change the effective area much and we can approximate the localization distribution behaving similarly to a delta distribution.
\end{itemize}
With these simplifications and with changing the order of $d\mathbf{\Omega}_v$ and $d\mathbf{\Omega}$ integrations in the denominators, we have a simplified likelihood ratio.
\begin{widetext}
\begin{multline}
\frac{P(\nu|GW,\mathcal{T})}{P(\nu|\mathcal{F})}
=\frac{1}{n_{\rm av}}
\times \int P_{\rm GW}(\mathbf{\Omega},r)  \omega Poi\left(1,\int \frac{A_{\rm eff}(\mathbf{\Omega},\epsilon)}{4\pi  r^2\epsilon^2} \frac{E_{\rm \nu tot}f_{\rm b}\mu}{\ln(\epsilon_{\rm max}/\epsilon_{\rm min})}d\epsilon\right)drd\mathbf{\Omega}
\\ \times \sum_{i=1}^N P_{\rm emp}(\epsilon_i,\mathbf{\Omega}_i)^{-1}\left(\frac{T_{\rm obs}}{\Delta t}-\frac{T_{\rm obs}|t_m-t_i|}{\Delta t ^2}\right)\frac{\epsilon_i^{-2}A_{\rm eff}(\mathbf{\Omega}_i,\epsilon_i)\int P_{\rm GW}(\mathbf{\Omega})P_{\nu}(\mathbf{\Omega}|\mathbf{\Omega}_i,\sigma_i)d\mathbf{\Omega}}{\int [\epsilon_\nu'^{-2}\int P_{\rm GW}(\mathbf{\Omega})A_{\rm eff}(\mathbf{\Omega},\epsilon_\nu')d\mathbf{\Omega}]d\epsilon_\nu'} 
\\
+\frac{2}{n_{\rm av}^2}
\times \int P_{\rm GW}(\mathbf{\Omega},r)  \omega Poi\left(2,\int \frac{A_{\rm eff}(\mathbf{\Omega},\epsilon)}{4\pi  r^2\epsilon^2} \frac{E_{\rm \nu tot}f_{\rm b}\mu}{\ln(\epsilon_{\rm max}/\epsilon_{\rm min})}d\epsilon\right)drd\mathbf{\Omega}
\\ \times \sum_{i=1}^{N-1} \sum_{j=i+1}^N \bigg[ P_{\rm emp}(\epsilon_i,\mathbf{\Omega}_i)^{-1}\left(\frac{T_{\rm obs}}{\Delta t}-\frac{T_{\rm obs}|t_m-t_i|}{\Delta t ^2}\right)P_{\rm emp}(\epsilon_j,\mathbf{\Omega}_j)^{-1}\left(\frac{T_{\rm obs}}{\Delta t}-\frac{T_{\rm obs}|t_m-t_j|}{\Delta t ^2}\right) \\ \times \frac{\epsilon_i^{-2}\epsilon_j^{-2}A_{\rm eff}(\mathbf{\Omega}_i,\epsilon_i)A_{\rm eff}(\mathbf{\Omega}_j,\epsilon_j)\int P_{\rm GW}(\mathbf{\Omega}) P_{\nu}(\mathbf{\Omega}|\mathbf{\Omega}_i,\sigma_i)P_{\nu}(\mathbf{\Omega}|\mathbf{\Omega}_j,\sigma_j)d\mathbf{\Omega}}{\int [\epsilon_\nu'^{-2}\epsilon_\nu''^{-2}\int P_{\rm GW}(\mathbf{\Omega})A_{\rm eff}(\mathbf{\Omega},\epsilon_\nu')A_{\rm eff}(\mathbf{\Omega},\epsilon_\nu'')d\mathbf{\Omega}]d\epsilon_\nu'd\epsilon_\nu''} \bigg]
\\
+\int P_{\rm GW}(\mathbf{\Omega},r) \left\{1- \omega\left[1-\exp\left(-\int \frac{A_{\rm eff}(\mathbf{\Omega},\epsilon)}{4\pi  r^2\epsilon^2} \frac{E_{\rm \nu tot}f_{\rm b}\mu}{\ln(\epsilon_{\rm max}/\epsilon_{\rm min})}d\epsilon\right)\right]\right\}drd\mathbf{\Omega}
\end{multline}
\end{widetext}

\section{Applying to the real-time follow-up of coincident high-energy neutrinos with the subthreshold GW triggers during O4a}
\label{sec:data}
Having established the formalism, we proceed with applying it to data. The IceCube Collaboration conducts real-time searches for coincident high-energy neutrinos with the GW triggers that are publicly shared in low-latency by the LIGO Scientific Collaboration, the Virgo Collaboration, and the KAGRA Collaboration through the Gravitational-Wave Candidate Event Database (GraceDB)\footnote{\url{https://gracedb.ligo.org/}}. The searches use IceCube's data sample called Gamma-ray Follow-Up (GFU)~\cite{2016JInst..1111009I,Kintscher_2016,Blaufuss:20199c} that is based on muon and antimuon tracks with which the direction of the original neutrino can be reconstructed within $\lesssim 1^\circ$. Starting from the fourth observing run of LIGO, Virgo, and KAGRA detectors (O4)~\cite{2020LRR....23....3A}, one of these searches, the LLAMA pipeline~\cite{countryman2019lowlatency}, conducts the search $\pm 500$~s around the GW triggers~\citep{2011APh....35....1B} including the subthreshold GW triggers as well which can have false-alarm rates up to 2/day. Their results are publicly available at \url{https://roc.icecube.wisc.edu/public/LvkNuTrackSearch/}. The searches output a frequentist $p$-value for the high-energy neutrino coincidence for each public GW trigger. The localization and detection time information for the coincident neutrinos, which produce a coincidence with a $p$-value of 0.1 or less, is shared publicly. On the other hand, the energy proxies for them are not shared publicly by the IceCube Collaboration. Due to the lack of the public availability of this one property, in order to proceed with the mathematical expression that we derived in the previous section we arbitrarily assign a range of possible energies within the sensitive energy band of IceCube (\{1, 10, 30, 100\}~TeV) to the released candidate coincident neutrinos and find corresponding new $p_{\rm astro}$ values for them. We limit ourselves to the coincidences that happened in the first half of O4 (O4a) as the second half is still ongoing. In O4a, there were a total of 1030 GW triggers that were analyzed for high-energy neutrino coincidences. Due to the undisclosed energies of neutrinos and in order to focus on the neutrino coincidences that may have actually produced a substantial increase in $p_{\rm astro}$, we restrict ourselves to coincidences that have a coincidence $p$-value of at most 0.01. We further concentrate on the subthreshold GW triggers to have meaningful and interesting improvements. These selections leave 11 GW event candidates that have a promising high-energy neutrino coincidence. Their neutrino coincidences do not involve overlapping two or more neutrinos. Hence, we do not calculate the multineutrino term in the example calculation, albeit it is possible and mathematically defined if the need arises from the data.

For our emission model, following Ref \cite{Kimura_2018} and typical short gamma-ray burst characteristics~\cite{2014ARA&A..52...43B}, we consider a beaming factor $f_{\rm b}\sim 100$; optimistic, moderate and pessimistic total neutrino emission energies of $E_{\rm \nu tot}\sim10^{49}$, $10^{48}$ and $10^{47}$ erg within the energy range of 0.5-500~TeV. We note that we consider the detection of neutrinos and antineutrinos only in muon flavor, which are expected to have a third of the total flux~\cite{ATHAR_2006,Pakvasa_2008}. Consequently, we take $\mu=1/3$. We estimated the fraction of GW detections from homogeneously distributed\footnote{Ignoring the cosmological evolution.} and isotropically oriented compact binaries that will point jets to us with $f_{\rm b}\sim 100$ as $\omega\approx 0.034$. This estimate simply assumes a detection based on matched-filter signal-to-noise ratio and the inclination-dependent GW emission from compact binaries. A more extensive investigation of this effect is left for future work.

We obtained the empirical estimation of the background characteristics of detected neutrinos and the effective area of IceCube via the public data release of the all-sky point source IceCube data between the years 2008-2018\footnote{\url{https://icecube.wisc.edu/data-releases/2021/01/all-sky-point-source-icecube-data-years-2008-2018/}}\cite{2021arXiv210109836I}. The average number of neutrinos expected within the 1000~s search window was taken as 6.4 \cite{PhysRevD.100.083017,Abbasi_2023}.

\begin{figure}
    \centering
    \includegraphics[width=\columnwidth]{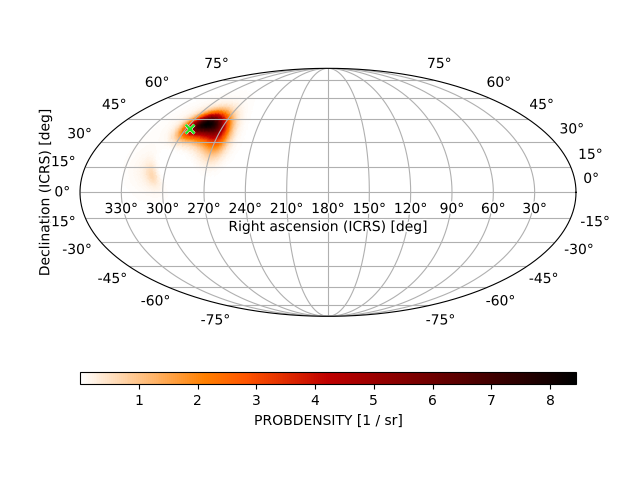}
    \includegraphics[width=\columnwidth]{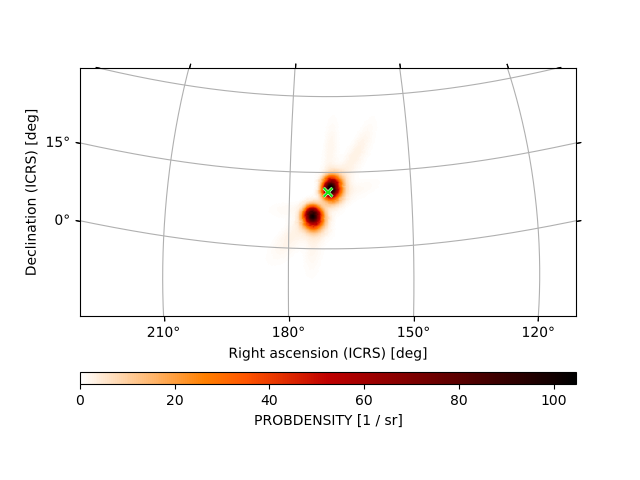}
    \caption{Example events: localizations of the coincident neutrino triggers (green crosses) and the GW triggers. Orange gradients represent the probability density for the localization of the GW events according to the color-bars. Top: S230812aj, Bottom: S231018cb}
    \label{fig:coincs}
\end{figure}

The $\mathcal{P}_{\rm astro}^{\rm MM}$ values for the 11 subthreshold GW events are shown in Table \ref{table:res}. It is observed, according to expectations, that in order to have a significant increase in the $p_{\rm astro}$ values, one needs highly energetic coincident neutrino which diverges from the expected background characteristics. For example, for coincident neutrino energy of 100~TeV and a total emission energy of $10^{48}$~erg, the $p_{\rm astro}$ of the trigger S230812aj can rise from 7.7\% to 86\% whereas for the optimistic $10^{49}$~erg it reaches 97\%. Similarly, the $\mathcal{P}_{\rm astro}^{\rm MM}$ of the trigger S231018cb can exceed 98\% and 90\% respectively for $10^{49}$ and $10^{48}$~erg while initially being only 16\%. For it, having a mildy energetic 30~TeV neutrino is enough to elevate its $p_{\rm astro}$ over 91\% and 55\%, respectively. Moreover, a 100~TeV neutrino can elevate its $p_{\rm astro}$ over 52\% even with the pessimistic total emission energy assumption of $10^{47}$~erg. Consequently, such coincidences would make these GW triggers confident detections. Figure \ref{fig:coincs} shows the localization of the coincident neutrinos of these two triggers on the sky. In addition, in the most optimistic combination (100~TeV and $10^{49}$~erg), the $p_{\rm astro}$ value of the trigger S231106y, which is originally less than 1\%, can be elevated over 60\%. 

\begin{table}[]
\centering
\begin{tabular}{|l|l|l|l|l|}
\hline GW trigger
& 1~TeV   & 10~TeV  & 30~TeV & 100~TeV \\ \hline
\begin{tabular}[c]{@{}l@{}}S230619bg\\ original = 0.0288\end{tabular} & 
\begin{tabular}[c]{@{}l@{}}0.0285\\(0.0288) \\ {[}0.0288{]}  \end{tabular}& 
\begin{tabular}[c]{@{}l@{}}0.0475\\(0.0325) \\ {[}0.0292{]}  \end{tabular} & 
\begin{tabular}[c]{@{}l@{}}0.298\\(0.0955)\\ {[}0.0369{]}   \end{tabular} & 
\begin{tabular}[c]{@{}l@{}}0.848\\(0.524) \\ {[}0.131{]}   \end{tabular} \\ \hline

\begin{tabular}[c]{@{}l@{}}S230628aj\\ original = 0.694\end{tabular}  & 
\begin{tabular}[c]{@{}l@{}}0.693 \\(0.694) \\ {[}0.694{]}   \end{tabular}  & 
\begin{tabular}[c]{@{}l@{}}0.716  \\(0.697) \\ {[}0.694{]}   \end{tabular} & 
\begin{tabular}[c]{@{}l@{}}0.882\\(0.742) \\ {[}0.700{]}  \end{tabular} & 
\begin{tabular}[c]{@{}l@{}}0.979 \\(0.883) \\ {[}0.738{]}   \end{tabular}  \\ \hline

\begin{tabular}[c]{@{}l@{}}S230701z\\ original = 0.144\end{tabular}   & 
\begin{tabular}[c]{@{}l@{}}0.144 \\(0.144) \\ {[}0.144{]}   \end{tabular}  & 
\begin{tabular}[c]{@{}l@{}}0.152\\(0.145) \\ {[}0.144{]}   \end{tabular}   & 
\begin{tabular}[c]{@{}l@{}}0.254\\(0.158) \\ {[}0.146{]}  \end{tabular} & 
\begin{tabular}[c]{@{}l@{}}0.727 \\(0.312) \\ {[}0.165{]}   \end{tabular}  \\ \hline

\begin{tabular}[c]{@{}l@{}}S230726b\\ original = 0.305\end{tabular}   & 
\begin{tabular}[c]{@{}l@{}}0.305 \\(0.305)\\ {[}0.305{]}    \end{tabular}  & 
\begin{tabular}[c]{@{}l@{}}0.349 \\(0.311) \\ {[}0.306{]}   \end{tabular}  & 
\begin{tabular}[c]{@{}l@{}}0.674\\(0.397) \\ {[}0.316{]}  \end{tabular} & 
\begin{tabular}[c]{@{}l@{}}0.954\\(0.758)\\ {[}0.421{]}    \end{tabular}   \\ \hline

\begin{tabular}[c]{@{}l@{}}S230812aj\\ original = 0.0774\end{tabular} & 
\begin{tabular}[c]{@{}l@{}}0.0771 \\(0.0773) \\ {[}0.0774{]}   \end{tabular} & 
\begin{tabular}[c]{@{}l@{}}0.238  \\(0.109) \\ {[}0.0811{]}   \end{tabular} & 
\begin{tabular}[c]{@{}l@{}}0.742\\(0.358)\\ {[}0.121{]}   \end{tabular} & 
\begin{tabular}[c]{@{}l@{}}0.973      \\(0.859) \\ {[}0.433{]}   \end{tabular} \\ \hline

\begin{tabular}[c]{@{}l@{}}S230908b\\ original = 0.0108\end{tabular}  & 
\begin{tabular}[c]{@{}l@{}}0.0107 \\(0.0108) \\ {[}0.0108{]}   \end{tabular} & 
\begin{tabular}[c]{@{}l@{}}0.0118  \\(0.0110) \\ {[}0.0108{]}   \end{tabular}& 
\begin{tabular}[c]{@{}l@{}}0.0309\\(0.0139)\\ {[}0.0111{]}   \end{tabular} & 
\begin{tabular}[c]{@{}l@{}}0.235 \\(0.0530) \\ {[}0.0155{]}   \end{tabular}  \\ \hline

\begin{tabular}[c]{@{}l@{}}S231018cb\\ original = 0.157\end{tabular}  & 
\begin{tabular}[c]{@{}l@{}}0.158 \\(0.157) \\ {[}0.157{]}   \end{tabular}  & 
\begin{tabular}[c]{@{}l@{}}0.408  \\(0.193) \\ {[}0.160{]}   \end{tabular} & 
\begin{tabular}[c]{@{}l@{}}0.911\\(0.552) \\ {[}0.226{]}  \end{tabular} & 
\begin{tabular}[c]{@{}l@{}}0.988 \\(0.901) \\ {[}0.521{]}   \end{tabular}  \\ \hline

\begin{tabular}[c]{@{}l@{}}S231025a\\ original = 0.588\end{tabular}   & 
\begin{tabular}[c]{@{}l@{}}0.588 \\(0.588) \\ {[}0.588{]}   \end{tabular}  & 
\begin{tabular}[c]{@{}l@{}}0.651  \\(0.604) \\ {[}0.590{]}   \end{tabular} & 
\begin{tabular}[c]{@{}l@{}}0.869\\(0.717) \\ {[}0.609{]}  \end{tabular} & 
\begin{tabular}[c]{@{}l@{}}0.957 \\(0.854) \\ {[}0.662{]}   \end{tabular}  \\ \hline

\begin{tabular}[c]{@{}l@{}}S231106y\\ original = 0.00356\end{tabular} & 
\begin{tabular}[c]{@{}l@{}}0.00356\\(0.00356)\\ {[}0.00356{]}    \end{tabular} & 
\begin{tabular}[c]{@{}l@{}}0.0168 \\(0.00529)\\ {[}0.00375{]}    \end{tabular}& 
\begin{tabular}[c]{@{}l@{}}0.104\\(0.0179) \\ {[}0.00508{]}  \end{tabular} & 
\begin{tabular}[c]{@{}l@{}}0.607  \\(0.169) \\ {[}0.0239{]}   \end{tabular} \\ \hline

\begin{tabular}[c]{@{}l@{}}S231205c\\ original = 0.00168\end{tabular} & 
\begin{tabular}[c]{@{}l@{}}0.00168 \\(0.00168)\\ {[}0.00168{]}    \end{tabular}& 
\begin{tabular}[c]{@{}l@{}}0.00539 \\(0.00220)\\ {[}0.00174{]}    \end{tabular}& 
\begin{tabular}[c]{@{}l@{}}0.0596\\(0.0102) \\ {[}0.00259{]}  \end{tabular} & 
\begin{tabular}[c]{@{}l@{}}0.462  \\(0.108) \\ {[}0.0141{]}   \end{tabular} \\ \hline

\begin{tabular}[c]{@{}l@{}}S231215i\\ original = 0.00126\end{tabular} & 
\begin{tabular}[c]{@{}l@{}}0.00126 \\(0.00126)\\ {[}0.00126{]}    \end{tabular}& 
\begin{tabular}[c]{@{}l@{}}0.00147 \\(0.00129) \\ {[}0.00127{]}   \end{tabular}& 
\begin{tabular}[c]{@{}l@{}}0.00383\\(0.00160) \\ {[}0.00130{]}  \end{tabular} & 
\begin{tabular}[c]{@{}l@{}}0.0375 \\(0.00545) \\ {[}0.00171{]}   \end{tabular} \\ \hline
\end{tabular}
\caption{$\mathcal{P}_{\rm astro}^{\rm MM}$ values of the subthreshold GW triggers coincident with the neutrinos which have $p$-value~$\leq 0.01$ from the real-time coincidence search, for a range of possible undisclosed neutrino energies, assuming a total neutrino emission energy of $10^{49}$~erg ($10^{48}$~erg) [$10^{47}$~erg]. We provide 3 significant figures according to the precision of the public data.}
\label{table:res}
\end{table}

\section{Discussion}

After providing an exact formalism and a concrete and precise example for describing how multimessenger association can elevate the significance of an individual detection by assuming a joint emission model and detailing it for a joint GW and high-energy neutrino case, we provided relevant application examples using public data. Here, we discuss several points further elucidating the method's application:
\begin{itemize}
    \item The $\mathcal{P}_{\rm astro}^{\rm MM}$ value in the formalism still refers to the probability of the \emph{originally} detected signal being astrophysical, that is the GW signal in the above application. Although information from other detectors is used, the $\mathcal{P}_{\rm astro}^{\rm MM}$ values do not quantify the significance of the coincidence between the detections of different detectors directly. One could only say that a ``significant" or ``nice" coincidence of different messengers; for example, where the different messengers have high overlap in their localizations, close detection times, and individual astrophysical characteristics as if they have come from the same astrophysical source, should increase the $p_{\rm astro}$ under the assumption of joint emission.
    \item A crucial indispensability of using the information of additional messengers accurately is the assumption of a joint emission model. This is related to the fact that a model-independent optimal solution for the multiple-messenger search problem does not exist~\cite{2021ApJ...908..216V}; thus, we need to know \emph{how} the messengers are related in order to have accurate statistics. Such problems necessitate Bayesian formalisms for accurate solutions, which thus, we apply Bayesian priors to describe the joint emission. However, there is \emph{no assumption} on joint detection. Neither coinciding different messengers are assumed to originate from the same source, nor is a conclusion about their joint detection made. Therefore, as long as one has trust in a model, the updating scheme of $p_{\rm astro}$ can be reliably used. 
    \item The most robust use of $\mathcal{P}_{\rm astro}^{\rm MM}$ would be with already observed messenger coincidences such as GW and short gamma-ray bursts from binary neutron star mergers for which we have a better understanding of the emission model~\citep{Abbott_2017_gwgrb}. In that case, the $\nu$ messenger in Eq. \eqref{eq:elevated} can be taken as the short gamma-ray burst.
    \item Can we use this updating scheme for not-yet-observed multimessenger coincidences? The application in Sec. \ref{sec:data} is such a case. Although there have been searches for joint detection of GW and high-energy neutrinos, no sufficiently significant joint detection could have been made~\cite{Albert_2019,Aartsen_2020,2022icrc.confE.950V,Abbasi_2023}. One should be cautious in such a case about what they assume. In such a situation, or in general, of course, the assumptions on the joint emissions should be consistent with the existing observations. For example, the searches for the joint detection of GW and high-energy neutrinos have yielded 90\% upper limits on the isotropically equivalent energies in the range of $\mathcal{O}(10^{51}-10^{55})$~erg~\cite{Abbasi_2023}. In our assumed model, the isotropically equivalent emission energy within the emission cone would be $10^{51}$~erg for the optimistic case. The lowest reported upper limit and our optimistic case barely coincide, whereas our realistic and pessimistic models' energies lie orders of magnitude below the reported observational upper limits. Consequently, the theoretical prediction and observational constraints are consistent with each other. Of course, if there are observationally estimated emission energies rather than upper limits, the trust in the priors of the model and in the results will be higher. 
    \item Do we need any kind of trials correction? One might reason that a high number of coincident background secondary messengers (e.g., neutrinos in the above example) would require trials correction to compensate for too many ``nice" false coincidences; however, that is not the case here. That effect is totally and accurately compensated by the factors proportional to the negative powers of $n_{\rm av}$, the average number of background neutrinos, in the calculation. 
    \\
    Another reason for a trials correction might be a high number of original triggers (e.g., too many subthreshold GW events). In such a scenario, there could be a substantial number of events that have ``nice" but false coincidences. Would this not produce an overall excess significance? It would not as long as priors are accurate; because the calculation also accounts for such probabilities and increases the $p_{\rm astro}$ values accordingly. If a false coincidence is more probable for a given detection, for example, a detection with a poor sky localization, then a possible $p_{\rm astro}$ increase would be less due to lower probability density ($P_{\rm GW}(\mathbf{\Omega}$)) at the coinciding location in the sky. Although less, there will still be $p_{\rm astro}$ increases for false coincidences. However, for such an event, the improvement would also be small for the true coincidences, and on average, the expected change is balanced. Moreover, the $p_{\rm astro}$ increase is positively correlated with the expected number of true multimessenger detections. If that expectation is low, then the significance improvements will again be suppressed. However, the $p_{\rm astro}$ improvement calculation naturally expects to have a certain fraction of multimessenger coincidences to be true according to the assumed emission model. An overall excess or lack of significance could be caused by an inaccurate emission model; but not by the number of original triggers. This is the reason why we demonstrated how the $p_{\rm astro}$ changes depend on the assumed model by considering three different emission energies in our example.
    \item Finally, we mention an extra use case of $\mathcal{P}_{\rm astro}^{\rm MM}$. Although it may not be the best way to do it, $\mathcal{P}_{\rm astro}^{\rm MM}$ can be used to test the consistency of the models used in it with the observations. The use of extra messenger information should provide a more precise estimation of the probability of astrophysical origin. However, as long as $\mathcal{P}_{\rm astro}^{\rm MM}$ and $p_{\rm astro}$ are both accurate, they should be consistent with each other. This consistency can be checked with a test that has the same logic as a $\chi^2$ test. However, since $p_{\rm astro}$ for each event defines a Bernoulli distribution instead of a normal distribution, the $\chi^2$ statistic would not be optimal here. Instead we can use either of the following two equivalent test statistics which test for the inconsistency of two Bernoulli distributions:
    \begin{subequations}
    \begin{equation}
        \prod_{i=1}^n\left(p_{\rm astro}[i] \mathcal{P}_{\rm astro}^{\rm MM}[i]+(1-p_{\rm astro}[i] )(1- \mathcal{P}_{\rm astro}^{\rm MM}[i])\right)^{-1}
    \end{equation}
    or
    \begin{equation}
        \sum_{i=1}^n-\ln\left(p_{\rm astro}[i] \mathcal{P}_{\rm astro}^{\rm MM}[i]+(1-p_{\rm astro}[i] )(1- \mathcal{P}_{\rm astro}^{\rm MM}[i])\right)
    \end{equation} 
    \end{subequations}
    where $i$ runs over the events in the dataset. One can construct a frequentist test with these test statistics by simulating multimessenger emissions according to the assumed emission model and $p_{\rm astro}$ values, calculate $\mathcal{P}_{\rm astro}^{\rm MM}$ for the simulated events and emissions, and construct a background distribution for the test statistic. Then, the test statistic value of the real observations can be compared with the background distribution to find whether we can reject the assumed emission model with high confidence. Of course this scheme assumes $p_{\rm astro}$ values are accurate. Any inaccuracy in their calculation can also imply an inconsistency. Nevertheless, a better way of testing emission models would be constructing a test based on a coincidence test statistic used in the coincidence searches (e.g.,~\cite{PhysRevD.100.083017,2021ApJ...908..216V}) as the root cause of an evident inconsistency would be the significantly different number of coincidences.
\end{itemize}

\section{Conclusion}
\label{sec:conc}
Multimessenger associations can be used to increase the significance of individual detections. These significance increases can lead to new discoveries through the power of synergistic astrophysics where multimessenger observations are seamlessly folded into a single multifaceted virtual observatory via model-dependent optimal statistics. In this paper, we describe the exact way of employing such associations to update the Bayesian posterior probability of having an astrophysical origin for individual signals, namely $p_{\rm astro}$, by assuming a joint emission model. We demonstrated  the mathematical approach through a concrete example, the prominent case of GW and high-energy neutrino coincidences. We laid out the formulation, and assuming a joint emission model we applied our formalism to an interesting subset of the public data from the high-energy neutrino follow-up of subthreshold GWs. Due to the nonpublic energy information on the released neutrino candidates, we assumed a range of possible energies for them and then computed the $\mathcal{P}_{\rm astro}^{\rm MM}$ values. We found that especially for energetic coincident neutrino detections, the improvement can be substantial and can elevate the GW detection candidates into confident detection status. Triggers with an initial $p_{\rm astro}$ of $\sim 0.1$ can become confident detections with $\mathcal{P}_{\rm astro}^{\rm MM}$ of $\sim 0.9$. Being a Bayesian method, our formulation is inevitably dependent on some of the underlying assumptions. In this case, the assumptions are the joint emission model we assumed. The model we assumed was a beamed emission with $f_{\rm b}\sim 100$, within a time window of 1000~s with an emission spectrum $\epsilon^{-2}$. We used three different total emission energy values ($E_{\rm \nu tot}\sim 10^{49}$, $10^{48}$, $10^{47}$~erg) to show how our results depend on it. The changes in $p_{\rm astro}$ values would decrease if detection of the secondary messenger gets harder, e.g., with a lower $E_{\rm \nu tot}$, and the final and initial values converge as expected. Nevertheless, the formalism and method presented allow the incorporation of any model that opens the way to the computation of model-dependent $\mathcal{P}_{\rm astro}^{\rm MM}$ value families. Finally, the changes can also cause the $p_{\rm astro}$ values to decrease in the case of no or not sufficiently significant coincidences. However, due to the high probability of having a nondetection, at least in our demonstration, such decreases are minuscule. 

With ongoing and future multimessenger follow-ups, the formalism we described here can be used to increase the number of detections from individual detectors without requiring any hardware improvement, thus increasing their efficiency and the impact of the heavy investment of our societies. Consequently, the frequency and quality of follow-ups can also increase, which in turn increases the chance of synergistic multimessenger discovery while decreasing the cost of individual follow-up efforts. Although we demonstrated the method on GW and high-energy neutrino coincidences, it can be used for different messengers and can be scaled for an arbitrary number of messengers and messenger combinations, e.g., for joint analyses of gamma-ray, x-ray, ultraviolet, optical, infrared, radio, neutrino, cosmic ray, and GW emissions. We stress once again that as a Bayesian method, the improvements are reliable for the assumed emission model. Models for not-yet-observed coincidences may be less reliable and should be used with caution. To avoid overestimating potential improvements, reliable or conservative emission models should be used. However, this model dependency can be used for testing the models. The $\mathcal{P}_{\rm astro}^{\rm MM}$ quantity is a key output of the emerging synergistic astrophysics field, where the optimal use of statistics allows multimessenger observations to reach their potential.

\section*{Acknowledgments}
This document was reviewed by the Publication Committee of the IceCube Collaboration. We thank Markus Ahlers for this review. We also thank Klas Hultqvist, Ali Kheirandish and Matthias Vereecken for their feedbacks and examinations on the idea. D.V. is grateful to The Scientific and Technological Research Council of T\"urkiye for their support through the grant No. 123C484. The Columbia Experimental Gravity group is especially thankful for the generous support from the National Science Foundation under the grant No. PHY-2207937. We extend our sincere gratitude to Heidelberg University, Columbia University in the City of New York, and Middle East Technical University for their generous support.
\bibliography{refs}
\end{document}